\documentclass[12pt]{article}
\usepackage{amsfonts,amsmath,epsfig,graphicx,amssymb}
\usepackage{amsmath,epsfig,graphicx,amssymb}
\usepackage[margin=1in]{geometry}
\usepackage{amsmath}
\usepackage{amsfonts}
\usepackage{amssymb}
\usepackage{makeidx}
\usepackage{graphicx}
\usepackage[utf8]{inputenc}
\usepackage[T1]{fontenc}
\newcommand{\beq}{\begin{equation}}
\newcommand{\eeq}{\end{equation}}
\newcommand{\ben}{\begin{eqnarray}}
\newcommand{\een}{\end{eqnarray}}
\date{}
\begin{document}
\title{Unveiling A Hidden Classical-Quantum Link}
\author{Partha Ghose\footnote{partha.ghose@gmail.com} \\
Tagore Centre for Natural Sciences and Philosophy,\\ Rabindra Tirtha, New Town, Kolkata 700156, India}
\maketitle
\begin{abstract}
The conceptual divide between classical physics and quantum mechanics has not been satisfactorily bridged as yet. The purpose of this paper is to show that such a bridge exists naturally in the Green-Wolf complex scalar representation of electromagnetic fields and its extension to massive fields. The quantum mechanical theory of radiation that follows from the the Green-Wolf representation is applied to the cosmic microwave background radiation (CMBR) regarded as a universal medium, and the implications are explored.
\end{abstract}
 
\begin{center}{\Large {\bf Part I: Electrodynamics}}
\end{center}

\section{Complex Scalar Representation of Classical Electrodynamics}
Green and Wolf have shown that classical electromagnetic fields in vacuum can be rigorously derived from a single {\em complex} scalar potential \cite{green, wolf, rom}. The Lagrangian density is
\beq
{\cal{L}}_\gamma = \xi \left(\partial_\mu\psi^*(x)\partial^\mu\psi(x)\right) = \xi\left(\frac{1}{c^2}\dot{\psi}^*\dot{\psi} - \vec{\nabla}\psi^*.\vec{\nabla} \psi\right)\label{L1}
\eeq
where the fundamental constant $ \xi =\hbar c l_P$ (with $l_P$ the Planck constant) has been introduced to make $\psi^*\psi$ have the dimension $L^{-3}$, and $x := (t, \vec{x})$. By comparing with the conventional Lagrangian density of free electromagnetic fields in terms of the vector potential $\vec{A}$ which satisfies the subsidiary condition $\vec{\nabla}.\vec{A} = 0$,
\beq
{\cal{L}}_{em} =  -\frac{1}{4} F^{\mu\nu}F_{\mu\nu} = \frac{1}{2}\left[\frac{1}{c^2}\dot{\vec{A}}.\dot{\vec{A}} - (\vec{\nabla}\times \vec{A}).(\vec{\nabla}\times \vec{A})\right],
\eeq 
we find the correspondence
\ben
\frac{1}{c^2}\dot{\psi}^*\dot{\psi} &\widehat{=}& \frac{1}{2}\vec{E}.\vec{E},\,\,\,\,\vec{E} = -\frac{1}{c}\vec{\dot{A}} \\
\vec{\nabla}\psi^*.\vec{\nabla}\psi &\widehat{=}& \frac{1}{2} \vec{B}.\vec{B},\,\,\,\,\vec{B} = \vec{\nabla}\times \vec{A}.\label{B}
\een
To make this correspondence more precise, let $n_\mu$ satisfying $n^\mu n_\mu = -1$ be a space-like unit tangent to the wave wavefront at any point, and let
\beq
A_\mu = \sqrt{\xi}\, n_\mu \mathcal{R}e(\psi)
\eeq
so that $\partial^\mu A_\mu = \sqrt{\xi}\, n_\mu\partial^\mu\mathcal{R}e(\psi) = 0$, and let $A_0 = 0$ in a particular frame so that $\vec{\nabla}.\vec{A}= 0$. Then,
\ben
F_{\mu\nu} &=& \partial_\mu A_\nu - \partial_\nu A_\mu = \sqrt{2\xi}\, (\partial_\mu n_\nu - \partial_\nu n_\mu)\mathcal{R}e(\psi),\\
-E_i &=&F_{0i} = \partial_0 A_i = \sqrt{2\xi}\, \partial_0\mathcal{R}e(\psi) n_i,\\
B_i &=& \epsilon_{ijk}\partial_j A_k = \sqrt{2\xi}\,\epsilon_{ijk} \partial_j\mathcal{R}e(\psi)n_k.
\een
This specifies the correspondence completely.

The variational equation that follows from (\ref{L1}) is 
\beq
\left(\nabla^2 -\frac{1}{c^2}\frac{\partial^2}{\partial t^2}\right)\psi(x) = 0 \label{C}
\eeq
which is the classical wave equation of a complex massless potential. 
This equation is invariant under Lorentz transformations $x^{\mu\prime} = \Lambda^\mu_\nu x^{\nu}$, and a plane wave solution can be written in the form
\beq
\psi(x) = A_k e^{-ikx} = A_k e^{i(\vec{k}.\vec{x} - k_0x_0)} 
\eeq
with $k^2 = k_0^2$ where $k^2 =\vec{k}.\vec{k}$, $k = 2\pi/\lambda$. This is a classical wave of amplitude $A_k$, and hence need not be normalized. It follows from this form of $\psi(x)$ that
\ben
(\partial^{\prime 2}_0 - \nabla^2_{\vec{x}^\prime}) \psi^\prime(x^\prime) &=& (-ik_0^\prime\partial^\prime_0 -\nabla^2_{\vec{x}^\prime})\psi^\prime(x^\prime)\\
&=&  (\partial_0^2 - \nabla^2_{\vec{x}})\psi(x)\\
&=& (-ik_0\partial_0 -\nabla^2_{\vec{x}})\psi(x) = 0.
\een
Therefore, using the definition $k_0 = \omega(k)/c$, one obtains the Lorentz invariant equation
\beq
i\dot{\psi} = - \frac{1}{\omega(k)/c^2}\nabla^2\psi. \label{Q}
\eeq
Now, writing the nonrelativistic Schr\"{o}dinger equation in the form
\beq
i\dot{\psi} = - \frac{\hbar}{2m}\nabla^2 \psi
\eeq
and comparing with eqn (\ref{Q}), one finds a surprising `correspondence' between the two: the left hand sides (including the coefficient $i$) are identical, and the right hand sides differ only by the coefficient of $\nabla^2$. However, note that 
\beq
\frac{1}{\omega(k)/c^2} = \frac{\hbar}{\hbar\omega(k)/c^2} := \frac{\hbar}{2m^*},\,\,m^* = \frac{\hbar\omega(k)}{2c^2}
\eeq 
and so eqn (\ref{Q}) has the form
\beq
i\dot{\psi} = - \frac{\hbar}{2m^*}\nabla^2\psi \label{sch}
\eeq
with $m^*$ as an `effective mass' which transforms like $\omega$ under Lorentz transformations. This is therefore a relativistic Schr\"{o}dinger-like equation for a massless {\em particle} with an `effective mass' $m^*$. The transition from a classical wave function to a quantum wave function with a particle interpretation is brought about by the introduction of the Planck constant. A more detailed exposition of this point will follow, but to arrive at it, let us
first apply the remaining time derivative in (\ref{Q}) on $\psi$ to obtain the classical Helmholtz equation
\beq
\left(\nabla^2 + k^2\right)\psi = 0,\,\,\,\,k^2 = \frac{\omega^2}{c^2} \label{Helm}
\eeq
where $k$ is the wave number in vacuum (refractive index $n = 1$). Most interestingly, {\em this classical equation} (\ref{Helm}) {\em is derivable from the classical wave equation} (\ref{C}) {\em via the intermediate equation} (\ref{Q}) {\em which has the mathematical structure of the Schr\"{o}dinger equation}!

To see the crucial difference between equations (\ref{Q}) and (\ref{Helm}), consider a general solution of eqn (\ref{Q}) of the form  
\ben
\psi(x) &=& \sqrt{\rho(x)}\,{\rm exp}(i\phi(x)),\label{polar1}
\een
where $\rho(x)$ and $\phi(x)$ are real Lorentz scalar functions. Substituting this in eqn (\ref{Q}) and separating the real and imaginary parts, one obtains the coupled equations
\ben
\frac{\omega}{c^2}\frac{\partial \phi}{\partial t} + (\nabla \phi)^2 &=& \frac{\nabla^2 \sqrt{\rho (x)}}{\sqrt{\rho (x)}},\label{disp}\\
\frac{\partial \rho}{\partial t} + \nabla.(\nabla \phi\rho(x)) &=& 0. \label{cont1}
\een
Thus, eqn (\ref{Q}) admits such general solutions provided the above conditions are satisfied.
For monochromatic waves $\phi(x) = \vec{k}.\vec{x} - \omega t$, and hence
$\partial \phi/\partial t = -\omega$, $\vec{\nabla} \phi = \vec{k}$, and one gets
\beq
k^2 = \frac{\omega^2}{c^2} + \frac{\nabla^2 \sqrt{\rho (x)}}{\sqrt{\rho (x)}}. \label{disp2}
\eeq
However, substitution of the same solutions for $\psi$ in the classical Helmholtz equation (\ref{Helm}) and separation of the real and imaginary parts result in the constraint
\beq
\frac{\nabla^2 \sqrt{\rho (x)}}{\sqrt{\rho (x)}} = 0\label{qp}
\eeq
from the real part.
This shows that the additional $x$-dependent term in eqn (\ref{disp2}), which causes dispersion, vanishes in the classical case, ensuring that classical wave packets
\beq
\psi(x) = \int \psi(k) e^{i(\vec{k}.\vec{x} - \omega(k)t)} d^3k
\eeq
are non-dispersive. 

But {{\em there is no such constraint on a wave function that satisfies eqn} (\ref{Q}). This opens up the possibility of a {\em non-classical} wave mechanics based on eqn (\ref{Q}) for dispersive wave packets. Since eqn (\ref{Q}) is the same as eqn (\ref{sch}), let us see what the implications are of incorporating a fundamental unit of action $\hbar$ into it. Since the action is $S = \int {\cal L}_\gamma d^4x$, scaling the wave function $\psi$ by an arbitrary parameter $\lambda$ implies that $S$ scales by the factor $\lambda^2$. However, if the scale of the action is set by a fundamental constant $\hbar$, then it is no longer permissible to scale the wave function $\psi$ arbitrarily, which means it must be normalized. That in turn implies that $\psi^*\psi$ {\em can be interpreted as a probability density}. That is, indeed, Born's rule. The wave function $\psi$ can be normalized by requiring
\beq
\int \psi^* \psi d^3 x = 1.
\eeq

Writing $\phi = S/\hbar$, one can rewrite the solution (\ref{polar1}) in the form
\begin{eqnarray}
\psi(x) &=& \sqrt{\rho(x)}\,{\rm exp}(iS(x)/\hbar), \label{polar}
\end{eqnarray}
where both $\rho$ and $S$ are real functions. Let us consider the stationary cases for which $S(x) = W(\vec{x}) - Et$. Separating the real and imaginary parts and substituting in eqn.(\ref{sch}), one obtains the coupled equations
\ben
\frac{\partial S(x)}{\partial t} + \frac{(\nabla S(x))^2}{2m^*} + Q &=& 0\label{Hamq}
\een
with
\ben
H = \frac{(\nabla S(x))^2}{2m^*} + Q &=& \frac{(\nabla W(\vec{x}))^2}{2m^*} + Q, \label{H}\\
Q &=& - \frac{\hbar^2}{2m^*}\,\frac{\nabla^2 \sqrt{\rho(x)}}{\sqrt{\rho(x)}},\label{Q1}
\een
and
\begin{eqnarray}
\frac{\partial \rho}{\partial t} + \vec{\nabla}.(\rho \vec{\nabla} W) = 0.\label{conteq}
\end{eqnarray}
Eqn. (\ref{Hamq}) is the Hamilton-Jacobi equation in electrodynamics and Eqn. (\ref{conteq}) is a conservation law (essentially the Poynting theorem). $Q$ is known in the literature as the `quantum potential'. Eqn (\ref{Hamq}) shows that the evolution of the phase $\phi(x,t) = S(x,t)/\hbar$ is dependent on the real part of the wave function $\sqrt{\rho(x)}$. This is a special feature of quantum mechanics absent in classical wave theory in which condition (\ref{qp}) holds, making $Q$ vanish even though $\hbar \neq 0$.

Using the relation $S/\hbar = (W - Et)/\hbar = \phi = \vec{k}.\vec{x} - \omega t$ for an eigenstate of energy and momentum, one gets the familiar quantum mechanical results 
\beq
E = \hbar\omega, \vec{\nabla} W = \vec{p} = \hbar \vec{k}. \label{Ep}
\eeq
It now follows from eqns (\ref{Hamq}) and (\ref{H}) that 
\beq
E = \frac{\hbar^2 k^2}{2m^*} + Q = pc + Q.\label{zero}
\eeq
This is a scaled version of eqn (\ref{disp2}) and is a very significant result which shows that $Q$, the quantum potential, is the purely quantum mechanical energy {\em which vanishes by condition} (\ref{qp}) {\em in classical wave theory, independent of $\hbar$}.  

To give a concrete example of $Q$, one can consider the case of a photon in a $1D$ box of length $a$. The well known solution is $\psi(x) = \sqrt{\rho(x)}\sin kx, \,\,k = n\pi/a,\,\,n = 1,2, \cdots$. Hence,
\beq
Q_n = \frac{\hbar^2n^2\pi^2}{2m^* a^2} = \frac{n\hbar c \pi}{a} \label{Eo}
\eeq
The lowest energy level corresponds to $n = 1$, and $Q_1 = \hbar \pi c/a$ is the zero-point energy. Instead of a box one can consider other time independent potentials also. It is straightforward to see that for a harmonic oscillator potential $\frac{1}{2}\beta x^2$, for example, the zero-point energy is $\frac{1}{2}\hbar \omega_0,\,\,\omega_0 = \sqrt{\beta/m^*} = \sqrt{2\beta c^2/\hbar \omega}$. Thus, the zero-point energy depends on the shape of the confining potential. 

Before passing on to the next topic, it would be worthwhile noting that eqn (\ref{zero}) can be written as
\beq
H = E = pc + Q
\eeq
from which follows the Hamiltonian equations
\ben
\dot{p}_i &=& -\frac{\partial H}{\partial x_i} = -\frac{\partial Q}{\partial x_i},\label{Bohm}\\
\dot{x} &=& \frac{\partial H}{\partial p} = c.
\een
The first of these equations has the form of Bohm's equation for a nonrelativistic massive particle and may be interpreted as a relativistic generalization of it \cite{bohm}.

\begin{flushleft} {\em Commutation Relations}
\end{flushleft} 

Now consider the general operator equations 
\beq
[D_i,x_j] = -i\delta_{ij}\label{com}
\eeq
where $D_i = - i\partial_i$ is the displacement operator, which must hold in classical field theories. If one defines the momentum operators by $p_i = \hbar D_i$, this commutator can be written in the standard quantum mechanical form
\begin{equation}
[p_i,x_j] = -i\hbar\delta_{ij}.\label{qcom}
\end{equation}
It is usually argued that this commutator vanishes in the limit $\hbar\rightarrow 0$, the classical limit. The mathematics, however, shows that in the limit $\hbar\rightarrow 0$ what one actually gets is $0 = 0$. {\em The underlying non-commutative structure} (\ref{com}) {\em is independent of $\hbar$}. In classical theory the translation operator $D_i =-i\partial_i$ and $x_i$ do not commute. In quantum mechanics the operator $\hbar D_i$, interpreted as the momentum operator $p_i$, does not commute with the position operator $x_i$. Although the physical interpretations are different, the underlying mathematical structure is the same. The change in physics comes through the Planck constant $h$ which sets a new scale for action missing in the classical theory.
 
\begin{flushleft} {\em Classical and Quantum Waves Functions}
\end{flushleft} 

Finally, let us write $\psi(x) = \langle x|\psi\rangle$. Let $|\psi\rangle = \sum_i c_i|\psi\rangle_i$ where $|\psi\rangle_i$ form a complete basis in a Hilbert space. If one defines the operator $P_i = |\psi\rangle_i\langle\psi|$ and scale $|\psi\rangle$ by $\lambda$, $P_i^{\prime 2} = \lambda^2P_i$ and it cannot be idempotent, i.e. it cannot be a projection operator. However, if $|\psi\rangle$ is normalized, $\sum_i|c_i|^2 = 1$ and $P_i$ is idempotent. Let us consider the pure state $\rho = |\psi\rangle\langle \psi|$. It satisfies the conditions $\rho^2 = \rho$, ${\rm Tr} \rho =1$. If an observable $\hat{O} = \sum_i a_i P_i$ with discrete and nondegenerate eigenvalues $a_i$ and projectors $P_i = |i\rangle\langle i|$ is measured on the system, then according to the L\"{u}ders rule \cite{lud} the state is updated to 
\beq
 \rho \rightarrow \rho^\prime_k = \frac{P_k\rho P_k}{{\rm Tr} P_k\rho} \label{luders}
\eeq
{\em on the condition that the result $a_k$ was obtained}. However, if one considers the total state {\em without selection or reading of individual results}, the state transforms to 
\beq
\hat{\rho} = \sum_k p_k\rho^\prime_k = \sum_k P_k\rho P_k 
\eeq
where $p_k = {\rm Tr} P_k\rho$ is the probability weight of the state $\rho^\prime_k$ in the full ensemble \cite{busch}. This is the von Neumann rule. The L\"{u}ders rule clarifies its meaning and applicability. 

It is clear from these discussions that the fundamental differences between classical and quantum wave functions arise from two features. First, classical wave functions satisfy condition (\ref{qp}) but quantum wave functions do not. Second, the Planck constant $\hbar$ sets a new scale which requires the quantum wave function to be normalized, giving rise to discreteness in energy and momentum, projection operators and the special nature of projective measurements. The classical wave function can be arbitrarily scaled and has no such features. This scaling enables the amplitude and hence the intensity of classical waves to be varied arbitrarily. That freedom is not available to quantum wave functions which are normalized. 

Interestingly, {\em wave functions that satisfy the quantum mechanical equation} (\ref{sch}) {\em are readily derivable from the classical wave equation} (\ref{C}) for monochromatic plane waves which are eigenstates of energy and momentum. 

In support of this one can cite the well known experiments of Aspect and his group \cite{gr1} who have shown very clearly that classical light pulses remain classical no matter how weak (low intensity) they are made by inserting neutral density filters---they always produce classically expected coincident counts on a beam splitter. The idea that a sufficiently low intensity light pulse cannot contain more than one photon and hence must be quantum mechanical, is contradicted by experiments. To observe the particle or quantum nature of light, one has to produce single photon light pulses (or squeezed states) which, when of sufficiently low flux, produce `anti-coincidence on a beam splitter', the unambiguous signature of particle-like behaviour. Hence, a state of light is either classical or quantum depending on how it is prepared or produced---{\em there is no transition from one to the other}. A coherent state of quantum light is the nearest one can get to classical light, but it is essentially quantum in nature. There is thus a {\em contextuality and complementarity between classical and quantum light: the full nature of light can only be comprehended by taking into account the mutually exclusive methods of preparing these two forms of light}. And this is readily understood in terms of the theory outlined above.

An important feature of time-independent quantum mechanical wave functions is their single-valuedness. In nonrelativistic quantum mechanics this follows from the ellipticity of the Schr\"{o}dinger equation and the fact that Euclidean space is simply-connected \cite{rie}. Since the Helmholtz equation is also elliptic, classical wave functions that are solutions of this equation must also be single-valued in simply connected spaces.

\begin{flushleft}{\em Helicity}
\end{flushleft}

Let us next see how the helicity of electromagnetic radiation is described in the Green-Wolf scalar theory. Following Wolf \cite{wolf} we write 
\begin{equation}
\psi(\vec{x},t) = \psi_+(\vec{x},t) + \psi_-(\vec{x},t) 
\end{equation}
with
\begin{eqnarray}
\psi_+(\vec{x},t) &=& \int_{-\infty}^0 \Psi(\vec{x}, \omega){\rm exp}(-i\omega t) d\omega = \int_0^\infty\Psi(\vec{x}, -\omega){\rm exp}(i\omega t) d\omega,\\
\psi_-(\vec{x},t) &=& \int_{0}^\infty \Psi(\vec{x}, \omega){\rm exp}(-i\omega t) d\omega 
\end{eqnarray}
satisfying the relativistic Schr\"{o}dinger-like equation, but $\psi$ being complex, they are not complex conjugates of each other in general. They have been termed {\em partial waves} by Wolf who has shown that on time averaging, $\psi_+$ and $\psi_-$ are incoherent and represent two independent circularly polarized components of light with helicity $\pm 1$, i.e.
\begin{equation}
\int_V \psi^*_{\pm}\hat{\lambda} \psi_{\pm} = \pm 1
\end{equation}
where $\hat{\lambda} = \frac{{\bf \sigma}.{\bf p}}{|{\bf \sigma}||{\bf p}|}$. It follows from this that
\begin{equation}
\int_V \psi^*\hat{\lambda} \psi = 0
\end{equation}
which shows that the convection current
\beq
\vec{j} = -\frac{i\hbar}{2}\left[\psi^* \vec{\nabla}\psi -\vec{\nabla}\psi^* \psi\right]\label{cur}
\eeq
does not have any helicity. However, the currents 
\ben
\vec{j}_{+} &=& -\frac{i\hbar}{2}\left[\psi_+^* \vec{\nabla}\psi_+ -\vec{\nabla}\psi_+^* \psi_+\right]\\ 
\vec{j}_{-} &=& -\frac{i\hbar}{2}\left[\psi_-^* \vec{\nabla}\psi_- -\vec{\nabla}\psi_-^* \psi_-\right]\nonumber\\
\een
carry $\pm 1$ helicities. 

Finally, the current $\vec{j}$ satisfies the continuity equation
\beq
\partial_\mu j^\mu = \frac{\partial \rho}{\partial t} + \vec{\nabla}.\vec{j} = 0
\eeq
where $j_0 =\rho = \hbar k_0|\psi|^2 >0$.

Notice that the time component of the conserved current $j^\mu = \psi^*\partial^\mu\psi - \partial^\mu\psi^*\psi$ associate with the classical wave equation (\ref{C}) is not always positive definite due to the second time derivative in the equation requiring the choice of two initial conditions. Consequently, $\rho = cj^0 = \dot{\psi}^*\psi - \psi^*\dot{\psi}$ cannot be interpreted as a probability density. This was the historical reason for the Pauli-Weisskopf second quantization of the Klein-Gordon equation \cite{pw}. By contrast, the time component of the conserved Schr\"{o}dinger-like current $j_\mu = (\vec{j}, \rho)$ is positive definite and can be interpreted as a probability density. Hence, the Pauli-Weisskopf second quantization is not mandated.

\begin{flushleft}{\em Entanglement in Classical Optics}
\end{flushleft}

Finally, let us consider entanglement. It is by now well known that entanglement occurs in classical optics \cite{sp, gh, eb, ai}. The reason is now clear--the classical and quantum mechanical wave functions are mathematically related. 

Suppose there is a bipartite classical state $|\psi\rangle_{AB} \in H_A\otimes H_B$ where $H_A$ and $H_B$ are two Hilbert spaces. Then, according to the Schmidt decomposition theorem (which dates back to 1907 \cite{schm} and is pre-quantum) it is always possible to express this state as
\beq
|\pi\rangle_{AB} = \sum_{i=0}^{d-1} \lambda_i |i\rangle_A|i\rangle_B \label{ent}
\eeq
where $\lambda_i$ are real and strictly positive, $\sum_i\lambda_i^2 = 1$, and $\{|i\rangle_A\}, \{|i\rangle_B\}$ are orthonormal bases in $H_1, H_2$ respectively. The Schmidt rank $d$ of a bipartite state is equal to the number of Schmidt coefficients $\lambda_i$ in its Schmidt decomposition and satisfies
\beq
d \leq {\rm min}\{{\rm dim}(H_A), {\rm dim}(H_B)\} 
\eeq 
If the Schmidt rank $d > 1$, the state is entangled, i.e. it cannot be written as a product state.

In classical optics it is always possible to consider light of unit intensity without implying normalization in the quantum mechanical sense. Hence, the mathematical result (\ref{ent}) is equally applicable to classical and quantum mechanical optics. 

The two Hilbert spaces $H_A, H_B$ in the Schmidt decomposition have {\em disjoint} bases: $\{|i\rangle_A\}\cap \{|i\rangle_B\} = \emptyset$. There is, obviously, nothing in the mathematical theorem that tells us how the disjointness is to be physically realized.
For {\em intra-system} bipartite entanglement (i.e. entanglement between two different degrees of freedom of a single system), one can have, for example, path-polarization entanglement in classical optics and path-spin entanglement in quantum mechanics where the choice of paths (strictly speaking, disjoint spatial modes) is restricted to two. For {\em inter-system} entanglement (entanglement between two different systems) one can have polarization-polarization entanglement in both classical and quantum optics, the dimension of the Hilbert spaces being $2$ in both cases. In this case the spatial wave functions of the two systems remain in product form. So far only intra-system entanglement has been experimentally studied in classical optics \cite{ai}, but extension to inter-system entanglement is possible in principle. 

{\em It should be clear therefore that the mathematical structure of entanglement is fundamentally the same for classical and quantum mechanical radiation, though because of the normalization of the quantum states forced by the Planck constant, projective measurements play a role in quantum radiation that has no counterpart in classical radiation}.

\begin{flushleft}{\em Interaction with Matter}
\end{flushleft}

Before passing on to the implications, let us briefly consider the interaction of radiation with Dirac particles. The Lagrangian desnity is
\beq
{\cal L} = \bar{\Psi}\left(i\gamma^\mu\partial_\mu - mc/\hbar\right)\Psi + \partial_\mu\psi^*\partial^\mu\psi + e\bar{\Psi}\gamma^\mu\Psi\partial_\mu\psi
\eeq
which is invariant (to within total four-divergence terms) under the local gauge transformations $\Psi^\prime = e^{ie\theta}\Psi, \psi^\prime = \psi + \theta$ $(\theta$ real) with the restriction $\Box \theta$ = 0. 

\begin{center}{\Large {\bf Part II: Massive Electrodynamics}}
\end{center}
The Green-Wolf complex scalar representation of electromagnetic fields in vacuum turns out to be crucial in formulating a satisfactory theory of radiation encompassing both its classical and quantum aspects. We will now show that it can be extended to massive fields.

Following Part I, let us consider the Lagrangian density of a massive complex scalar field in vacuum,
\beq
{\cal{L}} = \hbar c l_P \left(\partial_\mu\psi^*\partial^\mu\psi - \mu^2\psi^*\psi\right)
\eeq
where $\mu$ is an arbitrary constant with the dimension of inverse length. Notice that it is possible to choose $\mu = l_P^{-1}$. The classical wave equation that follows from it is
\beq
\left(\nabla^2 -\frac{1}{c^2}\frac{\partial^2}{\partial t^2} - \mu^2 \right)\psi(x) = 0. \label{A}
\eeq
Let
\beq
\psi(x) = A_k e^{i(\vec{k}.\vec{x} - k_0x_0)}
\eeq
be a monochromatic plane wave solution with $k^2 + \mu^2 = k_0^2$.
Following the same procedure as in Part I and applying the operator $\partial_0$ once on $\psi$ results in the equation
\beq
i\frac{\partial \psi}{\partial t} = \left[-\frac{c}{k_0}\nabla^2 + \frac{\mu^2 c}{k_0}\right]\psi,\,\,\,\,k = \sqrt{\omega(k)^2/c^2 - \mu^2}, \label{clsch}
\eeq
which has the mathematical structure of the Schr\"{o}dinger equation with a $k_0$ dependent potential, and can indeed be written in the form 
\beq
i\frac{\partial \psi}{\partial t} =\left[-\frac{\hbar}{2m^*}\nabla^2 + V_0\right]\psi\label{rsch}
\eeq
where
\beq
m^* = \frac{\hbar k_0}{2c},\,\,\,\,V_0 = \frac{\mu^2 c}{k_0}.
\eeq 
The principal difference from the non-relativistic Schr\"{o}dinger equation is the occurrence of the effective mass $m^*$ which transforms like $k_0$ under Lorentz transformations. Writing $m^* = \gamma m_0$, $\gamma = (1 - v^2/c^2)^{-1/2}$ and ignoring terms of $\mathcal{O}(v^2/c^2)$, one obtains 
\begin{equation}
i\hbar\frac{\partial \psi}{\partial t} = \left[-\frac{\hbar^2}{2m_0}\nabla^2 + V^\prime_0\right]\psi,\,\,\,\,V^\prime_0 = \frac{\hbar^2\mu^2}{2m_0} = \frac{1}{2}m_0c^2,
\end{equation} 
which is the nonrelativistic Schr\"{o}dinger equation with a constant potential. This shows that eqn.(\ref{rsch}) {\em is the correct relativistic generalization of the Schr\"{o}dinger equation}. 

Applying the time derivative on $\psi$ in eqn (\ref{clsch}), one obtains the Helmholtz equation
\ben
\left(\nabla^2 + k^2\right)\psi &=& 0,\label{Helm2}\\
k^2 &=& k_0^2 - \mu^2 = \frac{\omega(k)^2}{c^2} - \mu^2 \label{Helm22} 
\een
where $k$ is the wave number in vacuum (refractive index $n = 1$). Most interestingly, {\em this classical equation} (\ref{Helm2}) {\em is derivable from the classical wave equation} (\ref{A}) {\em via the intermediate equation} (\ref{clsch}) {\em which has the mathematical structure of the Schr\"{o}dinger equation}!

Eqn (\ref{rsch}) ensures that the convection current
\begin{equation}
\vec{j} = -\frac{i\hbar}{2}\left[\psi^* \vec{\nabla}\psi -\vec{\nabla}\psi^* \psi\right]
\end{equation}
is conserved, 
\begin{equation}
\frac{\partial \rho}{\partial t} + \vec{\nabla}.\vec{j} = 0,\,\,\,\,\rho = \hbar k_0\psi^*\psi >0
\end{equation}
and its time component $\rho > 0$. {\em It is therefore possible to interpret $\psi^*\psi$ as the position probability density and $\vec{j}$ as the probability current density} when $\psi$ is normalized. Hence, eqn (\ref{rsch}) can be given a quantum mechanical particle interpretation.

The classical wave equation (\ref{A}) does not have this property because the time component of its conserved current $j^\mu = \psi^*\partial^\mu\psi - \partial^\mu\psi^*\psi$ is not always positive definite due to the second time derivative in the equation requiring the choice of two initial conditions. Consequently, $\rho = cj^0 = \dot{\psi}^*\psi - \psi^*\dot{\psi}$ cannot be interpreted as a probability density. This was the historical reason for the Pauli-Weisskopf second quantization of the Klein-Gordon equation which is no longer mandated.

Now consider a general solution of eqn (\ref{clsch}) in the polar form  
\ben
\psi(x) = \sqrt{\rho(x)}\,{\rm exp}(i\phi(x)),\label{polar21}
\een
where $\rho (x)$ and $\phi(x)$ are real functions. Substituting this in eqn (\ref{clsch}) and separating the real and imaginary parts, one obtains the coupled equations
\ben
\frac{k_0}{c}\frac{\partial \phi}{\partial t} + (\nabla \phi)^2 + \mu^2 &=& \frac{\nabla^2 \sqrt{\rho (x)}}{\sqrt{\rho (x)}},\label{dispa}\\
\frac{\partial \rho}{\partial t} + \vec{\nabla}.(\vec{\nabla} \phi\rho(x)) &=& 0 \label{cont21}.
\een
These are therefore the conditions that must hold for general solutions of eqn (\ref{clsch}).
For monochromatic solutions, $\phi(x) = \vec{k}.\vec{x} - k_0 ct$,
$\partial \phi/\partial t = -k_0 c$, $\vec{\nabla} \phi = \vec{k}$, and one gets
\ben
k^2 =k_0^2 &-& \mu^2 + \frac{\nabla^2 \sqrt{\rho (x)}}{\sqrt{\rho (x)}}\nonumber\\
&=& \frac{\omega(k)^2}{c^2} - \mu^2 + \frac{\nabla^2 \sqrt{\rho (x)}}{\sqrt{\rho (x)}} \label{disp22}
\een
Substitution of the same solutions in the classical Helmholtz equation (\ref{Helm2}) and separation of the real and imaginary parts result in the condition
\beq
\frac{\nabla^2 \sqrt{\rho (x)}}{\sqrt{\rho (x)}} = 0\label{qp2}
\eeq
from the real part.
This shows that the additional $x$-dependent term in eqn (\ref{disp22}), which causes dispersion, vanishes in classical theory, ensuring that classical wave packets are non-dispersive.  

But, as in the massless case studied in Part I, {\em there is no such restriction on a wave function satisfying eqn} (\ref{clsch}), which therefore forms the basis of a non-classical wave mechanics with dispersive wave packets. 

Eqn (\ref{rsch}) with $m^* = \hbar\omega/2c^2$ is the same as eqn (\ref{clsch}) except that the unit of the action $S$ is set to be $\hbar$. Hence it is no longer permissible to scale the wave function $\psi$ arbitrarily, which means it must be normalized and can be interpreted as a probability density. Other important consequences of normalization of the wave function have been discussed in Part I.

Since $\psi$ describes a massive field, there is a longitudinal component of the polarization vector in this case in addition to two transverse components.

\section{Quantum and Classical Particles}
Multiplying eqns (\ref{dispa}) and (\ref{cont21}) by the arbitrary unit of action $\eta$ and writing $\eta\phi = S$, we get
\ben
\frac{\partial S}{\partial t} + \frac{(\nabla S)^2}{2m} +  V_0 + {\cal Q} &=& 0,\label{hj}\\
 {\cal Q} = -\frac{\eta^2}{2m}\frac{\nabla^2 \sqrt{\rho (x)}}{\sqrt{\rho (x)}},\,\,\,\,V_0 &=& \frac{\eta\mu^2 c}{k_0},\,\,\,\,2m = \eta k_0/c.\nonumber
\een
and
\beq
\frac{\partial \rho}{\partial t} + \vec{\nabla}.(\vec{\nabla} S\rho) = 0.\label{cont2} 
\eeq
Eqn (\ref{hj}) is the Hamilton-Jacobi equation for a massive particle in a potential $V_0 + {\cal Q}$. For stationary eigenstates of energy and momentum one can set $S = W - Et$, $\vec{p} (= \gamma m\vec{v}) = \vec{\nabla S} = \vec{\nabla} W$. Then, 
\beq
H = \frac{p^2}{2m} +  V_0 + {\cal Q}
\eeq
and hence
\ben
\dot{x}_i = \frac{\partial H}{\partial p_i} = \frac{p_i}{m},\\
\dot{p}_i = - \frac{\partial H}{\partial x_i} = -\frac{\partial {\cal Q}}{\partial x_i} \label{qN}
\een
Eqn (\ref{qN}) would the relativistic version of Bohm's equation for a massive particle in a quantum potential $Q$ if one were to identify $\eta$ with $\hbar$. It is the quantum potential that gives rise to interference of quantum particles \cite{ph}.

Notice that condition (\ref{qp2}) prevents dispersion and at the same time causes ${\cal Q}$, the term responsible for quantum mechanical coherence, to vanish. It is therefore a sufficient condition for Newton's equation to hold. Eqn (\ref{hj}) then takes the form
\ben
\frac{\partial S_{cl}}{\partial t} +  H &=& 0,\label{hjc}\\
H = \frac{(\nabla W_{cl})^2}{2m} +  V_0  &=& \frac{p^2}{2m} + V_0 = 0,
\een
It follows from this that
\ben
\dot{x}_i = \frac{\partial H}{\partial p_i} = \frac{p_i}{m},\\
\dot{p}_i = - \frac{\partial H}{\partial x_i} = 0. \label{N}
\een
The absence of interference indicates that there is no fixed phase relationship between different points of the wave amplitude. This follows from eqn (\ref{hjc}) which shows that the phase $\phi = S_{cl}/\eta$ is independent of $\sqrt{\rho(x)}$. Hence, one cannot write a coherent superposition $\sum_i c_i\psi_i$ of wave functions describing a classical particle. However, one can still write a density matrix:
\beq
\hat{\rho} = \sum_i |c_i|^2 |\psi\rangle_i\langle\psi|. \label{mix}
\eeq
A similar situation obtains in the Koopman-von Neumann wave theory of nonrelativistic classical mechanics in which the particle wave function satisfies the Liouville equation \cite{K, vN}.

All this shows that eqn (\ref{clsch}) for a wave function $\psi$ that satisfies condition (\ref{qp2}) is equivalent to Newton's equation of motion for a free massive particle. Hence, {\em the same equation, namely eqn (\ref{clsch}), holds for both quantum and classical mechanics of relativistic particles depending on whether or not the wave function satisfies a certain condition}. It describes quantum particles if the wave function is normalized and does not satisfy condition (\ref{qp2}), and classical particles if it is not normalized and satisfies condition (\ref{qp2}).

Eqns (\ref{qN}) and (\ref{N}) are second order differential equations in time and their solutions require two initial conditions specifying the position and velocity which can be varied independently. In quantum mechanics this is not permissible and solutions of eqn (\ref{qN}) require special care. There is no such restriction on eqn (\ref{N}) which is classical. Further, in a theory in which the classical and quantum aspects of a system are intrinsically linked, they have the same ontology, and hence the de Broglie-Bohm type of interpretation \cite{bohm} is a natural choice.  

\section{Measurements}
Quantum mechanics presumes classical measuring apparatus with which quantum systems interact. This has been a fundamental problem since the inception of quantum mechanics because the two systems appeared so disparate, the quantum system being described by a ray in a Hilbert space and the classical system by a point in phase space. The option of treating the measuring apparatus also as a quantum system gave rise to the measurement problem which refuses to go away. A new option is now available, namely the use of a wave function in a Hilbert space for the classical measuring apparatus.

Let us consider the case of an observation designed to measure some observable $\hat{P}$ of a stationary quantum system $S$ with wave function $\psi_S({\bf x},t)$. Let the stationary classical wave function of the apparatus $A$ be $\psi_A(y,t)$ where $y$ is the coordinate of the `pointer'. The initial state is a product state
\begin{equation}
\Psi^{SA}({\bf x}_0,y_0, 0) = \psi^S({\bf x}_0,0)\otimes \psi^A(y_0,0) = \psi^A(y_0,0)\sum_p c_p \psi^S_p({\bf x}_0,0)
\end{equation}
where $\hat{P}\psi^S_p({\bf x}) = p\psi^S_p({\bf x})$. This is a hybrid wave function. This kind of wave function was first introduced by Sudarshan \cite{sud}. 

Following von Neumann, let us assume that the measurement interaction is impulsive, and that during this impulsive interaction the free evolutions of the quantum particle and the classical apparatus can be ignored because the mass of the particle is very large and the mass of the apparatus (the massive particle) can always be chosen to be sufficiently large. If one sets $\hbar = 1$ for convenience, the evolution operator of the system takes the form
\ben
U &=& {\rm exp}(-i\hat{\Omega} t),\\
\hat{\Omega} &=& -g\hat{P}\hat{D}_y \label{vN}
\een
where $g$ is a suitable coupling strength and $\hat{D}_y = -i\partial/\partial y$ is the classical displacement operator corresponding to the coordinate $y$ of the apparatus. The form (\ref{vN}) of the measurement interaction has been chosen to be of the von Neumann type. Then, 
\beq
U\psi^A(y,t) = e^{-igp \hat{D}_y t} \psi^A(y_0,0) = e^{-y_p \frac{\partial}{\partial y}} \psi^A(y_0,0) =  \psi^A(y_0 - y_p),\,\,\,\,y_p = gpt 
\eeq
for every $p$ and for $t \leq\tau$, the measurement time which is assumed to be extremely short. For $t >\tau$ there is no further displacement of the pointer. Hence, in accordance with (\ref{mix}), the final stationary state is of the form
\begin{equation}
\hat{\rho}^{SA} = \sum_p|c_p|^2 |p\rangle^S\langle p||p\rangle^A\langle p|.  \label{ent2}
\end{equation}
Each pointer position is correlated with a particular outcome $p$ with probability $|c_p|^2$, the correlation being exact in the limits of both $g$ and the number of trials tending to infinity. 

The mixed state $\hat{\rho}^S$ of the quantum system $S$ alone after the measurement can be obtained by tracing $\hat{\rho}^{SA}$ over the apparatus states: 
\beq
\hat{\rho}^S = {\rm Tr}_A\hat{\rho}^{SA} = \sum_p|c_p|^2 |p\rangle^S\langle p| 
\eeq
which is formally the same as the standard von Neumann mixed density matrix but does not imply a process of collapse.  

Thus, we have a unified theory of classical and quantum systems (intrinsically relativistic) in which measurement does not occupy any special significance, and the two systems naturally share the same ontology.

\begin{center}{\Large {\bf Part III: Implications}}
\end{center}
In the previous parts a hidden mathematical link between quantum and classical radiation has been used to develop a relativistic quantum mechanical theory of radiation (as opposed to second quantized quantum electrodynamics). In this part I will explore the possible implications of treating the cosmic microwave background radiation (CMBR) as such a quantum mechanical system and a universal medium.

\begin{flushleft}{(ii) \em Quantum Mechanics of Blackbody Radiation and CMBR} 
\end{flushleft}

In order to have a proper quantum mechanical theory of blackbody radiation, it is necessary to generalize the single photon wave function considered in Part I to the many-photon case. Consider a state of $N$ photons placed in $A = \sum_sA^s$ states with occupation numbers ($p_0, p_1,\cdots p_M$), $p_r = \sum_s p_r^s$, $N = N_{\rm max}^s$, each photon being placed in one of the states $|p_j\rangle,\,j\neq 0$, i.e.
\begin{equation}
\langle x_1,x_2,\cdots,x_N|p_0,p_1,p_2,\cdots p_M\rangle = \Pi_{j=1}^N \psi_{p_j}(x_j)  
\end{equation}
where $x_j =(\vec{x}_j,t)$ are the coordinates of the particles. The wave function with the correct permutation symmetry is therefore given by
\begin{equation}
\psi(x_1,x_2,\cdots,x_N) = \frac{1}{\sqrt{W}}\sum_{P\in \Lambda(p_1,\cdots, p_M)} P\langle x_1,x_2,\cdots,x_N|p_0,p_1,p_2,\cdots,p_M\rangle \label{multi}
\end{equation}
where $\Lambda(p_1,\cdots, p_M) \in S_M$ is the set of all permutations of the $p_j$ involving different $p_js$. This reflects the fact that a photon in a state with a given occupation number $p_j$ can come from any  of the positions $x_i$. This is the required generalization of the single photon wave function.

To derive the Planck formula for blackbody radiation, one can then follow Bose's method of distributing photons in such quantum states, calculating the macroscopically defined probability $W$ of a state having all types of quanta, the Boltzmann entropy $S = k {\rm ln} W$, and maximising it subject to the constraint that the total energy $E = \sum_s N^s h\nu^s$ remains fixed (see the Appendix for details).  

Now, the cosmic microwave background radiation (CMBR) has a Planck spectrum to a high degree of accuracy and can therefore be treated as blackbody radiation. It has been shown in Part I that quantum mechanical radiation has zero-point energy. According to Bose's  derivation of the Planck law (see Appendix), there are $\sum_s p_0^s$ states in Planck radiation which have no photons, where
\beq
p_0^s = A^s(1 - e^{-h\nu^s/kT}),\,\,A^s =\frac{8\pi\nu^{s2}d\nu^s}{c^3}
\eeq
according to the results (\ref{x1}, \ref{x2}) in the Appendix. These numbers vanish if $h\nu^s = 0$. They are therefore `vacuum states' with energy. The vacuum energy density of Planckian radiation is therefore 
\ben
\rho_{vac} &=& \int_0^{\omega_c} \frac{8\pi \omega^2}{(2\pi)^3 c^3}\hbar\omega \left(1 - e^{-\hbar\omega/kT}\right)d\omega\nonumber\\ 
&\simeq& \frac{\hbar^2 \omega_c^5}{5\pi^2 c^3 kT}  ,\,\,\,\,\frac{\hbar\omega}{kT}\ll 1 \label{vac1}
\een
where $\omega_c$ is a cut-off frequency. 
Since the CMBR spectral intensity is exponentially damped at high frequencies, its vacuum energy density must also be cut-off beyond some frequency $\omega_c$. One must now estimate the cut-off frequency. This can be done, for example, from the anomalous magnetic moment of the electron on the assumption that it is entirely caused by the CMBR vacuum density.

\begin{flushleft}{(iii) \em Anomalous Magnetic Moment of the Electron and Dark Energy}
\end{flushleft}

According to the Dirac theory, the electron has a spin magnetic moment $\langle\mu\rangle = g\mu_B \langle\sigma\rangle/\hbar = \mu_B$ where $\mu_B = e\hbar/2m_e c$, $\langle\sigma\rangle = \hbar/2$ and $g=2$. When placed in the CMBR, it interacts with the external electromagnetic field, and its vertex function is given by
\beq
\Gamma^\mu = F_1(q^2)\gamma^\mu + F_2(q^2)\frac{i\sigma^{\mu\nu}q_\nu}{2m} 
\eeq
where $q^2 = q^\mu q_\mu$ is the momentum transfer, and empirically the two form factors are known to satisfy the conditions $F_1(0) =1$ and $F_2(0) = a_e = (g -2)/2$. A non-zero $a_e$ is called the `anomalous magnetic moment' of the electron because in the reigning paradigm of QED the assumption is that $F_2(0) = 0$ {\em in vacuum} and in the absence of loop corrections to the vertex function. The non-zero value of $a_e$ is then shown to arise from loop corrections to the vertex function which are divergent, but a method exists to extract unambiguous finite results from them. In the one-loop approximation, $a_e \simeq \alpha/2\pi$ which is very close to the observed value \cite{Schw}.

In the relativistic quantum mechanics of radiation developed in the previous paper, the anomalous magnetic moment of the electron can be related to the CMBR vacuum energy in a non-perturbative and phenomenological way. It has also been shown in the previous paper that the quantum potential $Q$ is the source of all forms of quantum mechanical energy of radiation, including its zero-point energy. Now, the term $\nabla\psi^*.\nabla \psi$ in the Green-Wolf Lagrangian density for radiation (\ref{L1}),
\beq
{\cal{L}}_\gamma = \partial_\mu\psi^*\partial^\mu\psi =\frac{1}{c^2}\dot{\psi}^*\dot{\psi} - \nabla\psi^*.\nabla \psi,
\eeq
can be written as
\ben
\nabla\psi^*.\nabla \psi &=& -\nabla^2 \psi\, \psi^* + {\bf \nabla}.({\bf \nabla}\psi\, \psi^*)\nonumber\\
&=& -\frac{\nabla^2 \sqrt{\rho}}{\sqrt{\rho}}\rho + \cdots = \frac{\omega}{\hbar c^2}Q\rho + \cdots 
\een
where $\cdots$ is a total 3-divergence term. According to the correspondence (\ref{B}),
this energy density is magnetic. There is therefore a constant effective magnetic field ${\bf B}$ in the CMBR vacuum with energy density $\frac{1}{8\pi\mu_0}{\bf B}.{\bf B} = \rho_{vac}$. Electrons located within CMBR therefore acquire a Larmor energy 
\beq
\hbar\omega_L = a_e\mu_B B = \xi\rho_{vac}V
\eeq
where $0 < \xi \leq 1$ is an unknown `energy transfer efficiency factor' which we take to be unity to illustrate the basic physics.  
Therefore, 
\ben
a_e &=& \frac{\rho_{vac}}{\mu_B}\frac{V}{B} = \left(\frac{\hbar^2\omega_c^5}{5\pi^2 c^3 kT}\right) \left(\frac{2m_e c}{e\hbar}\right)\left(\frac{V}{B}\right)\nonumber\\
&=& \frac{5.5\times 10^{-71}\omega_c^5\, {\rm erg/cm^3}}{9.274\times 10^{-21}\,{\rm erg/G}}\left(\frac{\rm cm^3}{\rm G} \right)
\een
One obtains $a_e = \alpha/2\pi \simeq 0.0011614$ for $\omega_c = 2.87$ GHz. For such a value of $\omega_c$, $\rho_{vac} \sim 10^{-23}{\rm erg/cm^3}$ which is well within the observed upper bound $< 10^{-6}$ ${\rm erg/cm^3}$. 

In quantum electrodynamics, on the other hand, the vacuum energy per normal mode is $\hbar\omega/2$ and the result is
\beq
\rho_{vac}^{QED} = \frac{\hbar \omega_c^4}{8\pi^2 c^3}
\eeq
which diverges because there is no natural cut-off in the theory. Assuming a Planck energy scale cut-off, $\rho_{vac}^{QED}$ is $\sim 10^{114}{\rm erg/cm}^3$ which is some $120$ orders of magnitude larger than the observational upper bound \cite{ rugh, peebles}. 

The fact that the vacuum energy of CMBR calculated from the relativistic c-number quantum `mechanics' of radiation developed in this paper is well within the observed upper bound on it, unlike the QED value, is therefore a point in its favour.   
 
\begin{flushleft}(iv) {\em CMBR and Spontaneous Emissions}
\end{flushleft}

The presence of CMBR in the universe is ubiquitous, and practically all the light in the universe is spontaneously emitted. Might there be a connection between the two? There might indeed be one. To see how, consider a 2-level atomic system with an energy gap $E_2 - E_1 = h\nu$, and let a monochromatic beam of radiation of frequency $\nu$ and number density $N^\nu$ of photons be incident on it. The probability of absorption of the incident radiation by the system will be proportional to $n_1 N^\nu h\nu$ where $n_1$ is the number density of the atoms in the ground state. In the absence of CMBR the probability of emission of a photon by the atom will be $n_2 N^\nu h\nu$ where $n_2$ is the number density of the atoms in the excited state. In the presence of CMBR $N^\nu$ will change to $N^\nu + A^\nu$ where $A^\nu = 8\pi\nu^2/c^3$ is the number density of CMBR states with the same frequency. Hence the emission probability will change to $n_2 (N^\nu + A^\nu) h\nu$. Consequently, in thermal equilibrium the condition
\beq
\frac{n_2(N^\nu + A^\nu)h\nu}{n_1 N^\nu h\nu} = \frac{g_2}{g_1} \label{E}
\eeq
must hold, where $g_1, g_2$ are the degeneracies (multiplicities) of the two levels \cite{bose2}. This is, in fact, just the Planck law since the Boltzmann distribution law $n_1g_2/n_2g_1 = e^{h\nu/kT}$ must also hold. The additional term $A^\nu$ in the numerator causes spontaneous emissions with the correct Einstein coefficient.

Contrast this with the corresponding Einstein equilibrium condition  
\beq
n_1B_{12}\rho(\nu) = n_2 \left(A_{21} + B_{21}\rho(\nu)\right)
\eeq
together with the Boltzmann law, from which the Planck law follows only by imposing the ad hoc conditions $A_{21}/B_{21} = A^\nu$ and $B_{12}/B_{21} = g_2/g_1$. 

Historically, the basic Dirac theory of QED which gave the first explanation of spontaneous emission dates back to 1927, i.e.  much before the discovery of CMBR. In QED the spontaneous emission term arises from the commutation relation $[a^\dagger_{\nu}, a_{\nu^{\prime}}] = \delta_{\nu \nu^{\prime}}$ where $a^\dagger_{\nu}$ and $a_{\nu}$ are creation and annihilation operators of photons of frequency $\nu$, which gives rise to divergences and to an unacceptably large vacuum energy density in the universe. The alternative simple theory presented here is based on the application of a relativistic `quantum mechanics' of radiation to CMBR.

\begin{flushleft}{(v) \em CMBR and the Casimir Effect}
\end{flushleft}

A typical example of the Casimir Effect is the tiny attraction between two uncharged conductive plates placed a few nanometers apart in a vacuum \cite{cas}. The magnitude of the effect depends on the shape of the plates or the confining region. The effect is believed to be caused by the plates changing the vacuum energy of the electromagnetic field between them \cite{cas2}. All calculations using QED turn out to be divergent, but there are methods of regularization which can deal with that.

The vacuum energy of CMBR may also contribute to this effect. To illustrate this in the simplest case, let us calculate the effect produced by a $1D$ box of length $a$. We know from eqn (\ref{vac1}) that the vacuum energy of CMBR is $\rho_{vac} = C\omega_c^5$  where $C$ is a constant, and $\omega_c = \pi c/a$, $a$ being the separation between the plates. Hence 
\beq
P =\frac{\partial \rho_{vac}}{\partial a} = -\frac{5C\pi c\omega_c^4}{a^2} = -\frac{\hbar^2 \pi^3 c^2}{kT}\frac{1}{a^6}\simeq -\frac{7.5\times 10^{-17}}{a^6}\,{\rm dyne/cm^2}
\eeq
where $C = \hbar^2/5\pi^2 c^3 kT$ and $T = 2.7 K$. The negative sign indicates an attraction between the plates. A pressure $|P| \sim 10^9 {\rm dyne/cm^2}$ would thus result for $a \sim 4\times  10^{-5} {\rm cm}$.

\section{Concluding Remarks}
That the Green-Wolf complex scalar representation of electromagnetic fields would reveal the much needed mathematical and conceptual link between the c-number `quantum mechanics' of radiation and the classical field theory of radiation comes as a surprise. It lays the mathematical and conceptual foundation of wave-particle duality originally discovered by Einstein in the energy fluctuations of Planck radiation. The fundamental role of the Planck constant in forcing the normalization of the wave function and hence the Born rule, becomes transparent. The classical time independent Helmholtz eqn (\ref{Helm}) is derivable from the time dependent classical wave equation (\ref{C}) through the intermediate equation (\ref{Q}) which has an essentially Schr\"{o}dinger-like structure. 

The function $\frac{\nabla^2\sqrt{\rho}}{\sqrt{\rho}}$ plays a fundamental role in the theory. Its presence or absence determines whether the theory is quantum mechanical or classical. The Helmholtz equation forces this term to vanish, ensuring dispersion free classical waves in vacuum. Its presence allows non-classical, or quantum mechanical, waves to disperse in vacuum. It also determines the functional form of the quantum potential $Q$ responsible for all quantum mechanical features like quantum coherence and quantized energy levels.
It is noteworthy that the quantum potential, a typical feature of nonrelativistic de Broglie-Bohm theory \cite{bohm}, emerges naturally in a relativistic theory.

The generalization to massive electrodynamics is straightforward and leads to a theory of classical massive particles, and hence of classical measuring devices, obeying the Schr\"{o}dinger equation with the supplementary condition  $\frac{\nabla^2\sqrt{\rho}}{\sqrt{\rho}} = 0$ on the wave amplitude, and hence a satisfactory theory of measurement that does not require a collapse postulate.

The generalization of the Green-Wolf complex scalar representation to massless classical Yang-Mills fields and their quantum mechanical theory is under investigation. That should be of great importance for the standard model of particle physics as well as for Einstein's gravitational equations which have close relationships with Yang-Mills equations \cite{ash, mas}.

In Part III the relativistic quantum mechanical theory of radiation developed in Part I has been applied to CMBR treated as a universal medium. The implications are: (i) a finite vacuum energy of CMBR which is consistent with the observational upper bound on the cosmological constant, (ii) a finite anomalous magnetic moment of the electron immersed in CMBR, consistent with its observed value, (iii) a finite Casimir Effect  of the right order of magnitude due to CMBR vacuum energy, and (iv) a natural explanation of spontaneous emission of photons from atomic and molecular systems immersed in CMBR. In second quantized electrodynamics (QED) these effects are due to vacuum fluctuations at zero temperature, and the results are divergent though they can be regularized.

\section{Acknowledgement}
I am grateful to A. K. Rajagopal and Partha Nandi for many helpful discussions. 
 
\vskip 0.2in
{\centerline{\bf Appendix: Bose's Derivation of the Planck Law}}
\vskip 0.2in
It will be quite instructive to follow Bose's original derivation of Planck's law \cite{bose} which is quite different from the accounts given in text books and is therefore unknown to most physicists. Let us consider a collection of quantum states with frequency lying between $\nu^s$ and $\nu^s + d\nu^s$ in a hohlraum of volume $V$. Assuming a spherically symmetric $V$, one essentially has stationary radiation in a 1D box. As shown by Bose \cite{bose}, the total number of such states per unit volume in the range $d\nu^s$ is 
\begin{equation}
A^s = 2\int d^3p^s/h^3 = \frac{8\pi\nu^{s2}d\nu^s}{c^3} \label{As}
\end{equation}
(using $p^s = h\nu^s/c$), the factor 2 being due to helicity $\pm 1$. This is therefore {\em the number of possible arrangements of a single photon}. As argued by Bose, all possible arrangements of the photons in these states will correspond to a state $|p_0^s,p_1^s,p_2^s,\cdots\rangle$ where $p_0^s$ is the number of empty states, $p_1^s$ the number of 1-photon states, $p_2^s$ the number of 2-photon states, etc. Treating states with a given occupation number as identical, the probability of a state is given by
\begin{equation}
W^s = \frac{A^s!}{p_0^s! p_1^s!p_2^s!...},
\end{equation}
the number of photons of type $\nu^s$ being $N^s = \sum_r r p^s_r$ and $A^s = \sum_r p_r^s$. 
The macroscopically defined probability of a state having all types of quanta is thus
\begin{equation}
W = \Pi_s W^s = \Pi_s \frac{A^s!}{p_0^s! p_1^s!p_2^s!...},
\end{equation}
the total number of photons of all types being $N = \sum_s N^s$. Taking the $p_r^s$ to be large, one has
\begin{equation}
\ln W = \sum_s A^s \ln A^s -\sum_s\sum_r p_r^s \ln p_r^s
\end{equation}
with 
\begin{equation}
A^s = \sum_r p_r^s.
\end{equation}
This must be maximized subject to the constraint
\begin{equation}
E = \sum_s N^s h\nu^s = {\rm constant}.
\end{equation}
Carrying out the variations, one gets
\begin{equation}
\sum_s\sum_r \delta p_r^s (1 + \ln p_r^s) + \frac{1}{\beta}\sum_s h\nu^s\sum_r r\delta p_r^s = 0.\label{varn}
\end{equation} 
It follows from this that
\begin{equation}
p_r^s = B^s e^{-\frac{r h\nu^s}{\beta}}.\label{x1}
\end{equation}
But, since
\begin{equation}
A^s = \sum_r B^s e^{-\frac{r h\nu^s}{\beta}} = B^s \left(1 - e^{-\frac{h\nu^s}{\beta}}\right)^{-1},
\end{equation}
we have
\begin{equation}
B^s = A^s\left(1 - e^{-\frac{h\nu^s}{\beta}}\right).\label{x2}
\end{equation}

Further,
\begin{eqnarray}
N^s = \sum_r r p_r^s &=& \sum_r A^s\left(1 - e^{-\frac{h\nu^s}{\beta}}\right) e^{-\frac{r h\nu^s}{\beta}}\\
&=& \frac{A^s e^{-\frac{h\nu^s}{\beta}}}{1 - e^{-\frac{h\nu^s}{\beta}}},
\end{eqnarray}
and using the result (\ref{As}), one gets
\begin{equation}
E = \frac{8\pi\nu^{s2}d\nu^s}{c^3} V h\nu^s\frac{e^{-\frac{h\nu^s}{\beta}}}{1 - e^{-\frac{h\nu^s}{\beta}}}.
\end{equation}
Now,
\begin{equation}
S = k\left[\frac{E}{\beta} - \sum_s A^s \ln \left(1 - e^{-\frac{h\nu^s}{\beta}}\right)\right].
\end{equation}
Hence, using the relation $\frac{\partial S}{\partial E} = \frac{1}{T}$, we have $\beta = kT$. Substituting this in the expression for $E$ and dropping the suffix $s$ from $\nu^s$, we finally get the Planck formula
\beq
\rho(\nu) d\nu = \frac{E}{V} = \frac{8\pi\nu^2}{c^3} \frac{h\nu}{e^{\frac{h\nu}{kT}} -1} d\nu.
\eeq
Notice that only energy conservation plays a role in the derivation of this formula, but not photon number conservation.

\end{document}